\documentclass[a4paper,12pt]{article}
\pdfoutput=1
\usepackage{amsmath}

\usepackage{hyperref}
\hypersetup{colorlinks,citecolor=blue,urlcolor=magenta}
\usepackage{doi}

\usepackage[style=ext-authoryear-comp,
sorting=nyt,
dashed=false, 
maxcitenames=2, 
maxbibnames=99, 
uniquelist=false,
uniquename=false,
giveninits=true, 
natbib, 
date=year 
]{biblatex}

\AtBeginRefsection{\GenRefcontextData{sorting=ynt}}
\AtEveryCite{\localrefcontext[sorting=ynt]}

\DeclareFieldFormat{pages}{#1} 
\renewbibmacro{in:}{\ifentrytype{article}{}{\printtext{\bibstring{in}\intitlepunct}}} 

\DeclareFieldFormat[article,inbook,incollection,inproceedings,patent,thesis,unpublished]{titlecase:title}{\MakeSentenceCase*{#1}} 

\usepackage[inline,shortlabels]{enumitem}
\setlist[enumerate,1]{label=(\roman*)}

\usepackage[margin = 1.25in]{geometry}
\usepackage{setspace}
\title{Can AI Refute Economic Theory?\\ Evidence from Beyond the Knowledge Cutoff}
\author{Alexis Akira Toda\thanks{Department of Economics, Emory University. Email: \href{mailto:alexis.akira.toda@emory.edu}{alexis.akira.toda@emory.edu}. I thank Emi Nakamura for helpful comments and suggestions, and I gratefully acknowledge financial support from the Zengin Foundation.}}

\begin{document}
\maketitle
	
\begin{abstract}
Can artificial intelligence (AI) refute economic theory? I document experiments in which I asked several AI models (Gemini, Refine, Claude, and ChatGPT) to check the correctness of four published papers in economic theory, each containing an error that I helped identify or correct. ChatGPT Pro performed best, occasionally constructing counterexamples and corrected proofs, while other models fared worse. However, no model located a true error without substantial human guidance, and data contamination complicates interpretation. I argue that a competent human paired with a frontier model can outperform current peer review, but AI cannot yet refute economic theory on its own.

\medskip

		
\medskip

\end{abstract}
	
\section{Introduction}

In a recent paper, \citet{PhamToda2026ECMA} claim that Proposition 1(c) of \citet{Tirole1985} ``Asset Bubbles and Overlapping Generations'' is mathematically incorrect by presenting a counterexample (see their Proposition 1). Furthermore, they restore the original Proposition 1(c) under additional assumptions and show that such assumptions are essential. It may come as a surprise that a highly cited paper written by a Nobel laureate escaped scrutiny for more than 40 years.\footnote{According to Google Scholar, \citet{Tirole1985} is cited more than 2{,}000 times as of May 2026.} After the publication of \citet{PhamToda2026ECMA}, several people asked me if we used artificial intelligence (AI) to find the error or write the proof. The answer is no---we found the error and wrote the proof on our own, although we did use AI for grammar checking, as none of us are native English speakers.

The recent advance of AI is remarkable. Last year, several AI models reportedly performed poorly on USA Math Olympiad problems \citep{Petrov2025}. Only one year later, OpenAI announced that one of its large language models (LLMs) disproved the Planar Unit Distance Problem posed by Paul Erd\H{o}s in 1946.\footnote{\url{https://openai.com/index/model-disproves-discrete-geometry-conjecture/}} Systematic evaluations now show that LLMs coupled with formal proof assistants can resolve previously open problems, including several open Erd\H{o}s problems \citep{Tsoukalas2026}. The mathematical capabilities of frontier models have clearly improved quickly. Given this progress, it is of interest to test whether AI can refute economic theory. In this paper, I document my experiments applying several AI models to \citet{Tirole1985} and other papers.

My experiment proceeds as follows. I analyze four published papers in economic theory, namely \citet{Tirole1985}, \citet{Kocherlakota1992}, \citet{MiaoWang2018}, and \citet{StachurskiToda2019JET}. I select these papers deliberately: each contains an error that I helped identify or correct, so I know the flaw and its resolution intimately and am well positioned to prompt the models. For each paper, I upload the document to several AI models (Gemini, Refine, Claude, and ChatGPT), ask whether the key result is correct, and then challenge the model iteratively, steering it toward the problematic part as needed. To guard against the possibility that a model simply retrieves a known error or correction rather than reasoning to it, I take particular care with \citet{Tirole1985}, whose correction was published only in May 2026 in \citet{PhamToda2026ECMA}.\footnote{However, their working paper version (\url{https://arxiv.org/abs/2507.12477}) was posted in July 2025.} For the experiment with ChatGPT Pro on this paper, I disabled both the memory and the web-search functions before running the analysis.

My main findings are as follows. Among the models tested, ChatGPT Pro performed best, in several cases identifying the relevant flaw from a single prompt and constructing valid counterexamples and corrected proofs; Claude was weaker on formal reasoning but stronger on economic interpretation; and Gemini performed worst. In the \citet{Tirole1985} experiment, no model located a genuine error without my guidance: the initial verdict was always that the result was correct, or essentially so, and the flaw surfaced only after I steered the conversation toward it.

Two caveats are in order when interpreting these findings. First, a reader who only skims the AI transcripts (available from links included in footnotes throughout the paper) might conclude that the strongest model, ChatGPT Pro, would have located the error in the papers on its own. That is not the case. Because every model began by endorsing the proof of \citet{Tirole1985}, simply feeding a paper into an AI model and asking whether the result is correct is unlikely to reveal a deep error without substantial human guidance from someone who already suspects where the problem lies. The model lowers the cost of checking an argument once a candidate flaw is identified, but it does not locate the flaw. Second, the claim that a model ``found'' an error must be read against its \emph{knowledge cutoff}, that is, the date beyond which it has not been trained on new data. Several of the papers I analyze, together with their corrections, have circulated for some time, so a model may have seen the relevant material during training. This is precisely the concern raised by the literature on benchmark data contamination, which shows that apparent reasoning performance is inflated when evaluation material overlaps with training data \citep{Roberts2023,Jiang2024,Xu2024,Zhang2025}. The \citet{Tirole1985} experiment is the cleanest on this score: with the correction published only in May 2026 and the model's memory and web search disabled when I ran the experiment on May 29, 2026, I am reasonably confident that the counterexample ChatGPT Pro produced was genuinely constructed rather than retrieved. Furthermore, the generated counterexample was different from \citet{PhamToda2026ECMA}. Taken together, the experiments suggest that a competent economist working with a frontier model can already outperform the status quo of refereeing, even though AI cannot yet refute economic theory on its own.

The rest of the paper is organized as follows. \S\ref{sec:tirole} documents my experiments with Gemini, Refine, Claude Opus, and ChatGPT Pro on \citet{Tirole1985}. \S\ref{sec:other} reports analogous experiments on three other papers, namely \citet{Kocherlakota1992}, \citet{MiaoWang2018}, and \citet{StachurskiToda2019JET}. \S\ref{sec:discuss} discusses the implications for the use of AI
in economic theory and for peer review, and concludes.

\section{The case of \texorpdfstring{\citet{Tirole1985}}{}}\label{sec:tirole}

As mentioned in the introduction, Proposition 1(c) of \citet{Tirole1985} is incorrect. As I am not very tech-savvy, I did not know how to use AI to check mathematical correctness. As a first step, I asked Google Gemini 3.5 Flash (which is free and built into the Google Chrome browser).

\begin{quote}
I want to check the mathematical correctness of an economics paper. The paper is in a PDF format. Can you explain to me how to do it using Gemini (or other free AI tools)?
\end{quote}

Gemini gave me some instructions and I followed them.\footnote{\url{https://gemini.google.com/share/5eaafdf936f1}}

\subsection{Analysis of Gemini 3.5 Flash}

After the initial prompts, I directly asked about the proposition in question.

\begin{quote}
Is the proof of Proposition 1(c) logically correct?
\end{quote}

To this prompt, Gemini replied that ``The proof of Proposition 1(c) in Tirole's (1985) paper is logically sound'' and explained the key steps, including cases (i)--(iii) for the asymptotic behavior of the interest rate. Although I knew case (ii) was problematic, instead of flagging this case specifically, I asked for more clarification.

\begin{quote}
In the paper, the possibility of a bubbleless equilibrium with $\bar{r} < 0$ is considered at the bottom of page 1522, where Tirole states ``Let us now show that if $\bar{r} < 0$, there exists no [bubbleless] equilibrium.'' As you mentioned, Tirole considers three exhaustive and mutually exclusive cases. In each case, as you mention, Tirole reasons that $f_t\to 0$ and hence $r_t\to \bar{r}$. But I don't understand why $f_t\to 0$ implies $r_t\to \bar{r}$.
\end{quote}

Gemini replied that ``your skepticism is completely justified. The leap from $f_t \to 0$ to $r_t \to \bar{r}$ isn't immediately obvious from the text, but it relies on the continuity and monotonic properties of the temporary equilibrium mapping'' and provided some more details. However, Gemini's analysis was flawed because it claimed global stability from local stability, so I further challenged it.

\begin{quote}
Thank you for the clarification, but I still don't understand the argument. On p.~1502 of the Tirole paper, he states ``We also make Diamond's stability assumption that the curves defined by $\{w = \Phi(r)\}$ and $\{r = \psi(w, 0)\}$ have a unique intersection and that the $\psi$ curve is strictly steeper (in absolute value) than the $\Phi$ curve at this intersection in the $(r, w)$ plane''. This is a local argument, but your explanation seems to rely on a global argument. In general, a local stability condition guarantees that the dynamical system converges near the steady state, but there is no guarantee when the state is far away.
\end{quote}

Gemini admitted that ``You are completely correct that Diamond's stability assumption, as stated on page 1502, is a local stability condition'', but then claimed that global convergence can be obtained due to the monotonicity property in the Diamond model. However, this reasoning was flawed, so I challenged further.

\begin{quote}
Thank you for the clarification, but I think there is still an issue. If the OLG model does not have dividends (what Tirole calls rents), then the dynamics is one-dimensional, so the uniqueness of the steady state and the monotonicity property guarantee global convergence. But with positive dividends (rents), the dynamics is two-dimensional, so I don't think the proof works as you suggest. Am I missing something?
\end{quote}

To this prompt, Gemini explained the proof of Proposition 1(c) step-by-step. I then directly flagged the leap of logic.

\begin{quote}
I am still uncomfortable with Tirole's (and your) logic. In the logic flow you describe, I agree until Step 3. But in Step 4, Tirole (or you) claim that because $f_t\to 0$, the system behaves asymptotically like the one-dimensional Diamond model, forcing $r_t\to \bar{r}$. I think there is a leap of logic here. Consider Case (ii), where $\{r_t\}$ is strictly greater than $n$. Then we know that equilibrium forces $f_t\to 0$. But if $r_t\to \infty$ simultaneously, capital $k_t$ will converge to 0. Such a possibility does not seem to be ruled out by Tirole's argument.
\end{quote}

To my challenge, Gemini basically claimed that the Inada condition $f'(0)=\infty$ prevents equilibrium paths converging to zero. I knew this argument was flawed, so I simply uploaded \citet{PhamToda2026ECMA}, after which Gemini admitted its mistake.

Overall, after each prompt, Gemini 3.5 Flash provided a plausible argument, but it was not well-founded.

\subsection{Analysis of Refine}

Next, I turned to Refine.\footnote{\url{https://www.refine.ink/}} The reason I used this tool is that, when \citet{PhamToda2026ECMA} was conditionally accepted at \emph{Econometrica}, the editor ran our paper through Refine and asked us to provide feedback to the report it generated. (The Refine report was sophisticated but contained incorrect statements.)

To use Refine's service, it costs \$50 per analysis. Upon uploading \citet{Tirole1985}, Refine flagged several issues, but none of them looked serious; I found the report nitpicking. However, I used the chat function to discuss further.\footnote{\url{https://www.refine.ink/app/session/17a36e9c-5189-441b-a96e-b55ce478467a}} After several exchanges similar to Gemini, Refine admitted that the proof of Proposition 1(c) was incomplete. Furthermore, when I asked it to construct a counterexample (with few hints), it successfully generated a counterexample based on the Cobb-Douglas production function, though several details were omitted.

Overall, although the initial Refine report was somewhat underwhelming, its chat function was impressive. However, I am not sure if I would regularly use it from a cost perspective.

\subsection{Analysis of Claude Opus~4.8}

Next, I turned to Anthropic's Claude Opus~4.8 Pro, which costs \$20 per month. I did not use the more expensive (\$100) Claude Opus~4.8 Max because the pricing information suggested that Max allows for more volume and speed, not functionalities.\footnote{\url{https://claude.com/pricing}} My discussion with Claude was similar to Gemini: initially, Claude claimed that Proposition 1(c) was correct, but after I challenged a few times, it admitted that there was a leap of logic.\footnote{\url{https://claude.ai/share/255203c1-1de8-44d3-9536-cd8575caf2bd}} When I asked Claude to generate a counterexample, it tried the Cobb-Douglas function and derived necessary conditions for what the counterexample would look like, but it failed to generate one. When I provided the functional form
\begin{equation}
    f(k)=Ak\log(1+1/k) \label{eq:f_PT}
\end{equation}
used in the counterexample in \citet{PhamToda2026ECMA}, Claude solved the model numerically and claimed that it obtained a counterexample. However, upon inspection, it did not satisfy one of the assumptions of Proposition 1(c), so the analysis was incomplete.

Overall, I was impressed with Claude's computational capabilities, but not necessarily with its mathematical reasoning.

\subsection{Analysis of ChatGPT}

Finally, I turned to OpenAI's ChatGPT. First, I tried the free version.\footnote{\url{https://chatgpt.com/share/6a18d3ff-a5dc-83ea-80ea-0b717740a47a}} It made some basic mathematical mistakes, for example, mixing up the convergence of a subsequence with that of a sequence. (See the part titled ``6. The precise hidden argument'', where it mistakenly concludes $r_{t_k+1}\to r^*$ after establishing $r_{t_k}\to r^*$.)

Next, I upgraded to ChatGPT 5.5 Pro, which costs \$100 per month. ChatGPT Pro's analysis was truly amazing.\footnote{\url{https://chatgpt.com/share/6a1a519e-e57c-832d-9b82-53816c2b5946}} Even though I only asked ``Can you check the mathematical correctness of the proof of Proposition 1(c)?'', ChatGPT Pro flagged several issues and replied ``correct result, correct core logic, but not a fully rigorous proof as printed unless one supplies the missing boundary/supremum argument in the $\bar{r}<0$ existence step''. I then pointed out the key issue.

\begin{quote}
You mentioned ``The no-bubbleless-equilibrium part is essentially correct'' but I have an issue here. You mentioned ``If $f_t\to 0$, then the economy asymptotically behaves like Diamond's rentless/bubbleless economy, so $r_t\to \bar{r}<0$''. In the paper, the possibility of a bubbleless equilibrium with $\bar{r} < 0$ is considered at the bottom of page 1522, where Tirole states ``Let us now show that if $\bar{r} < 0$, there exists no [bubbleless] equilibrium.'' Tirole considers three exhaustive and mutually exclusive cases. In each case, Tirole reasons that $f_t\to 0$ and hence $r_t\to \bar{r}$. But I don't understand why $f_t\to 0$ implies $r_t\to \bar{r}$.
\end{quote}

With only this prompt, ChatGPT Pro (correctly) raised an issue with case (ii) discussed with Gemini and (correctly) concluded that ``this case does not itself give monotonicity [\ldots] the sentence ``$f_t\to 0$, hence $r_t\to \bar{r}$'' is not mathematically justified as a standalone implication''. Upon asking for a counterexample, ChatGPT Pro tried the case with a utility function exhibiting constant relative risk aversion (CRRA) $\gamma>1$ and the production function
\begin{equation}
    f(k)=\frac{A}{\gamma}k[\log(1/k)]^\gamma. \label{eq:f_ChatGPT}
\end{equation}
This suggestion was truly remarkable, as the special case of \eqref{eq:f_ChatGPT} with $\gamma=1$ (Cobb-Douglas) becomes $f(k)=Ak\log(1/k)$, which behaves exactly the same way as the function \eqref{eq:f_PT} as $k\to 0$. After several prompts, ChatGPT Pro was able to provide a counterexample with a complete proof.

Overall, I was hugely impressed with the mathematical ability of ChatGPT Pro.

\section{Other papers}\label{sec:other}

After my experiment with \citet{Tirole1985}, to test the robustness of my findings, I decided to feed other papers into AI models. Before running the experiment, I set the following rules.
\begin{enumerate}
    \item The papers I will test are \citet{Kocherlakota1992}, \citet{MiaoWang2018}, and \citet{StachurskiToda2019JET}.
    \item The AI models I will test are Gemini 3.5 Flash, Claude Opus~4.8, and ChatGPT 5.5 Pro.\footnote{I exclude Refine because I did not want to pay \$50 each time, and it was no better than ChatGPT Pro based on the previous experiment.}
    \item For each paper, my initial prompt would be to check the mathematical correctness of the key result. After the initial prompt, I will challenge as needed.
\end{enumerate}

To provide some context, I picked these three papers because each paper includes an incorrect result and I was involved in correcting or pointing out the error in each case. For instance, the original proof of Proposition 4 of \citet{Kocherlakota1992} was incorrect but \citet{KocherlakotaToda2023JET} corrected it under the original assumption; \citet{MiaoWang2018} defined ``bubble'' inconsistently throughout the paper but \citet{HiranoToda2025EJW} proved the nonexistence of bubbles in their model according to the standard definition; the proof of Proposition 5 of \citet{StachurskiToda2019JET} was incorrect but \citet{StachurskiToda2020Corrigendum} corrected by strengthening the assumptions (in an economically harmless way). Because I understand these papers well, it puts me in a good position to prompt AI models. 

\subsection{The case of \texorpdfstring{\citet{Kocherlakota1992}}{}}

For each AI model, the initial prompt was as follows.

\begin{quote}
I have just uploaded the paper ``Bubbles and Constraints on Debt Accumulation'' published in Journal of Economic Theory. Can you check the mathematical correctness of the proof of Proposition 4?
\end{quote}

\paragraph{Gemini 3.5 Flash}

Initially, Gemini responded ``The logic presented in the proof is mathematically sound''.\footnote{\url{https://gemini.google.com/share/1c05411055f6}} Upon challenging, Gemini (incorrectly) claimed that the existence of a subsequence $\{t_n\}$ and positive constant $b>0$ such that $a_{t_n}-a_{t_n-1}>b$ is a logical consequence of the assumption that $\{a_t\}$ does not converge. When I further challenged that there could be a sequence such that $\{a_t\}$ does not converge but $a_t-a_{t-1}$ converges to zero, Gemini admitted its mistake and provided the (correct) counterexample $a_t=\sin(\sqrt{t})$. When I asked if there is a known correction to \citet{Kocherlakota1992}, Gemini was unable to find \citet{KocherlakotaToda2023JET}. When I provided the exact link of the corrigendum, Gemini hallucinated.

\paragraph{Claude Opus~4.8}

Initially, Claude responded ``The proof of Proposition 4 is essentially correct in its logic, but it relies on a step that is asserted rather than fully justified, and there's a notational slip''.\footnote{
\url{https://claude.ai/share/5d4768af-2d65-4f31-bf3a-c5cbcbc9dd5a}} However, Step 2 contained a flaw. Upon challenging, Claude provided a counterexample very similar to the one in Section 3 of \citet{KocherlakotaToda2023JET} as well as $a_t=\sin(\sqrt{t})$ (same as Gemini). When I asked if there is a known correction to \citet{Kocherlakota1992}, Claude was able to find \citet{KocherlakotaToda2023JET}.

\paragraph{ChatGPT 5.5 Pro}

Just with my initial prompt, ChatGPT Pro identified the error in the proof of Proposition 4, provided the (correct) counterexample $a_t=\sin(\log(t+1))$, and furthermore, correctly proved the original statement, claiming that ``A corrected proof is straightforward''.\footnote{\url{https://chatgpt.com/share/6a1cc5bc-0c58-8333-82cf-403872e82c12}} In fact, ChatGPT Pro's proof is more elegant than \citet{KocherlakotaToda2023JET}! When I asked if there is a known correction to \citet{Kocherlakota1992}, ChatGPT Pro was able to find \citet{KocherlakotaToda2023JET}.

\subsection{The case of \texorpdfstring{\citet{MiaoWang2018}}{}}

For each AI model, the initial prompt was as follows.

\begin{quote}
I have just uploaded the paper ``Asset Bubbles and Credit Constraints'' and its online appendix. The abstract claims ``We provide a theory of rational stock price bubbles in production economies with infinitely-lived agents''. The authors claim on p.~2595 that ``Some studies (e.g., Scheinkman and Weiss, 1986; Kocherlakota, 1992, 2008; Santos and Woodford, 1997; Hellwig and Lorenzoni, 2009) have found that infinite-horizon models of endowment economies with borrowing constraints can generate rational bubbles. Unlike this literature, our paper analyzes a production economy with stock price bubbles attached to productive firms.'' The papers cited here are all ``rational bubble'' models, where the asset price exceeds the fundamental value defined by the present discounted value of dividends. Now, based on the standard definition of rational bubbles, does the model of Miao and Wang generate bubbles?
\end{quote}

\paragraph{Gemini 3.5 Flash}

Initially, Gemini (incorrectly) responded ``Based on the standard definitions and the specific modeling choices presented in the paper, the model of Miao and Wang does generate rational asset bubbles''.\footnote{\url{https://gemini.google.com/share/ca5c671327d8}} In the Miao-Wang model, the stock price takes the form $V_t=Q_tK_t+B_t$. Gemini claimed that $Q_tK_t$ is the fundamental value and $B_t$ is the bubble. When I asked if the authors prove that $Q_tK_t$ is the fundamental value, Gemini said yes, and cited the transversality condition. When I pointed out a contradiction, Gemini found my website, in which I discuss these issues, as well as my published article \citep{HiranoToda2025EJW} and other papers.

\paragraph{Claude Opus~4.8}

Claude (correctly) responded ``Based on the standard definition of a rational bubble, the answer is no'', ``the honest characterization is the following. The model generates multiplicity of equilibria---a ``bubbly'' and a ``bubbleless'' steady state---driven by self-fulfilling beliefs [\ldots]. But labeling the high-value equilibrium's extra term a ``rational bubble'' in the Santos-Woodford sense is a strain'', and ``Whether that broadened notion deserves the name ``rational bubble'' is exactly the point of contention, and reasonable people in the field disagree''.\footnote{\url{https://claude.ai/share/65098d32-6af6-46d3-9760-92a5889733d8}} Upon asking to elaborate on ``reasonable people in the field disagree'', Claude conceded that it was a speculation, but then searched the web and (correctly) summarized the exchange of \citet{HiranoToda2025EJW}, \citet{MiaoWang2025response}, and \citet{HiranoToda2025EJW_rejoinder}.

\paragraph{ChatGPT 5.5 Pro}

Initially, ChatGPT Pro (correctly) responded ``no, not in the usual Santos-Woodford/asset-pricing sense'' and ``Calling that object a ``bubble'' is internally consistent with their terminology, but it is not the same object as the rational bubbles in the papers they cite''.\footnote{\url{https://chatgpt.com/share/6a1d011e-a504-8331-b445-6fd18b2d6ff8}} When I asked for clarity about two different notions of ``transversality condition'' (TVC) and whether Miao and Wang actually proved it, ChatGPT (correctly) answered ``They do not prove Equation (6) as an independent household optimality/no-Ponzi TVC inside the paper. They mostly state it, cite general references, and then use it''. When I pointed out language issues in the abstract and introduction, ChatGPT Pro provided the verdict ``This makes the abstract and literature framing problematic. When they contrast their paper with Scheinkman-Weiss, Kocherlakota, Santos-Woodford, Hellwig-Lorenzoni, and related ``rational bubble'' models, they naturally invite the reader to understand their result in the same rational-bubble sense. [\ldots] They market this mechanism as a rational stock-price bubble, even though, under the standard Santos-Woodford present-value definition, their imposed TVC rules out such bubbles.'' Once I enabled web search, ChatGPT Pro found the exchange of \citet{HiranoToda2025EJW}, \citet{MiaoWang2025response}, and \citet{HiranoToda2025EJW_rejoinder} and concluded that Hirano and Toda's ``critique is essentially the same as the one we have been discussing'' (i.e., the analysis of ChatGPT Pro).

\subsection{The case of \texorpdfstring{\citet{StachurskiToda2019JET}}{}}

For each AI model, the initial prompt was as follows.

\begin{quote}
I have just uploaded the paper ``An impossibility theorem for wealth in heterogeneous-agent models with limited heterogeneity'' published in Journal of Economic Theory. Can you check the mathematical correctness of the proof of Proposition 5?
\end{quote}

\paragraph{Gemini 3.1 Pro}

Being disappointed with Gemini 3.5 Flash, I broke my own rule and switched to Gemini 3.1 Pro, which is free but supposed to be stronger at math and code. Gemini initially responded ``Based on a careful review of the provided paper, the mathematical proof of Proposition 5 is rigorous and logically sound''.\footnote{\url{https://gemini.google.com/share/0273b68c747d}} Upon challenging issues on continuity, Gemini admitted that the proof had a gap and suggested ideas to fix it. It also found the correction by \citet{StachurskiToda2020Corrigendum}.

\paragraph{Claude Opus~4.8}

After a lengthy analysis, Claude responded ``The proof of Proposition 5 is mathematically correct''.\footnote{\url{https://claude.ai/share/db71e110-e439-45c1-9f2c-6aebbf1edc88}} Upon challenging issues on continuity, Claude admitted the gap but claimed the proof can be rescued. When asked if a correction exists, it found \citet{StachurskiToda2020Corrigendum}, though it was unable to access. When I uploaded the corrigendum, Claude flagged that the original proof had a further issue (pointed out in the corrigendum).

\paragraph{ChatGPT 5.5 Pro}

Just with my initial prompt, ChatGPT identified key gaps in the proof of Proposition 5 and concluded ``the proposition is very likely true, but the proof as written is not fully mathematically correct''.\footnote{\url{https://chatgpt.com/share/6a1d1879-0508-8333-b4d0-89c56f7d25fa}} When asked if a correction exists, it found \citet{StachurskiToda2020Corrigendum} and confirmed that the proof of their Proposition 5' is correct.

\section{Discussion and conclusion}\label{sec:discuss}

My experiments yield three broad conclusions.

\subsection{Model performance and necessity of human input}

On the mathematics, ChatGPT Pro was the clear winner. It was the only model that, with no more than my initial prompt, repeatedly flagged the relevant issues, and in two cases (\citealp{Kocherlakota1992}, and the existence step in \citealp{Tirole1985}) it constructed a valid counterexample and a more elegant corrected proof. Claude Opus~4.8 was less mathematically sophisticated, tending to assert rather than justify the crucial step, but it was stronger on judgment and interpretation. In the case of \citet{MiaoWang2018}, for instance, it correctly resisted the ``rational bubble'' framing and characterized the disagreement in the field. Gemini was the clear loser. Across papers, it initially endorsed incorrect proofs, defended them with plausible but unfounded arguments, and on one occasion hallucinated when asked for a specific reference.

This performance ranking should not be read as a claim that the best model is an autonomous error-finder. This is the most important caveat of the paper, and it is easy to miss from the AI transcripts alone. In every case regarding \citet{Tirole1985}, the model's first response was that the proof was correct. The error came to light only because I knew in
advance which part was problematic, and I steered the conversation accordingly. A reader who simply uploaded the paper and asked whether the proof was correct would have been told yes. The AI models are excellent at following a line of reasoning once it is pointed out, at performing the algebra, and, in the case of ChatGPT Pro, at completing a proof whose key idea has been identified. But AI
models did not, on their own, decide that some results in the paper were incorrect.

This observation is consistent with the broader evidence on LLMs and mathematical reasoning. LLM-generated proofs are known to contain subtle errors that cascade through an argument and that require costly expert review to catch \citep{Petrov2025}. The practical implication for an economist is the following. Running a paper through even the best available AI model and asking whether the result is correct will probably not surface a deep error without substantial input from a human who already suspects where the problem lies. The AI lowers the cost of \emph{checking} an argument once a candidate flaw is identified. It does not yet reliably substitute for the human judgment
that \emph{locates} the flaw in the first place.

\subsection{Implications for peer review}

If AI changes what referees can check, it is worth asking what the refereeing process is for. One may view the editorial decision after the peer review process as a hypothesis test in which the null is that the paper is correct. Based on my experience and the broader episode surrounding the papers I analyzed, the current peer review process suffers from both kinds of error familiar from statistical decision theory. A Type~I error (the incorrect rejection of a correct contribution) occurred when \emph{American Economic Review} rejected \citet{HiranoToda2025EJW}: the original authors (who obviously have a conflict of interest) recommended rejection, and some of the three independent referees dismissed the issue as semantic. Based on their reports, none of them seemed to have read the proofs. A Type~II error (the incorrect acceptance of an incorrect contribution) is what happened with \citet{Tirole1985}, \citet{Kocherlakota1992}, \citet{MiaoWang2018}, and \citet{StachurskiToda2019JET}, each of which contains a flawed result that survived refereeing, though it was eventually corrected or critiqued. Both false positives and false negatives are costly, and the current peer review system controls neither well.

These failures are symptoms of structural problems in peer review rather than isolated anecdotes. \citet{Siemroth2024} documents problems in the current economics peer review system including referee overreach and excessive revisions, strategic refereeing and conflicts of interest, prestige bias and other discrimination, and the noisy outcome of peer review. In addition, refereeing is a public good provided under weak incentives (such as small cash compensations or recognition from editors), so even well-meaning referees rationally economize on effort. \citet{Azar2006} argues that although individuals have incentives to improve their own efficiency, no one has both the power and the incentive to fix the system as a whole. On the point of effort, my experience suggests that few referees read proofs in economic theory.\footnote{As an example, consider \citet{StachurskiToda2019JET}. As I document in my blog post (\url{https://alexisakira.github.io/publications/2019-JET/}), this paper was rejected from \emph{American Economic Review: Insights} (three reports), \emph{Econometrica} (desk rejection upon consulting an external advisor), and \emph{Review of Economic Studies} (four reports plus an external advisor), and accepted as is at \emph{Journal of Economic Theory} (one report). Among 10 reviewers/advisors who evaluated the paper, none seemed to have read the proof.} In mathematics, similar concerns about unread or under-checked proofs have motivated the push toward computer-assisted formal verification, reflecting a recognition that human refereeing alone cannot reliably certify long or intricate arguments \citep{Hales2008}.

Against this backdrop, AI offers a genuine but qualified opportunity. Based on my experiments, my tentative view is that for evaluating the internal logic of economic theory, a decently competent person using a frontier model such as ChatGPT Pro will likely outperform a typical referee who, under current incentives, does not read the proofs. The emphasis belongs on \emph{a person using AI}: the human supplies domain knowledge and targeted skepticism while the model supplies patient checking, algebra, and counterexample search. This has implications for editorial policy. Although some journals prohibit reviewers from uploading submitted papers to AI models, such a prohibition is difficult to enforce and, in my view, already out of date. The legitimate concern behind it is confidentiality, since a model provider could in principle train on the uploaded text. But this concern is easily sidestepped: journals could simply ask authors to post their papers on a public repository such as arXiv before submission, after which any confidentiality has already been waived.

\subsection{Limitations and the contamination problem}

Several limitations qualify these conclusions. First, the exercise is based on a small, deliberately selected set of papers whose errors I happen to know intimately. This is a feature for prompting quality but a limitation for external validity, and my role as an author of several of the corrections may color the interpretation. Second, the comparison across models is not fully controlled. Pricing tiers, default settings, and tool access differ, and the models are moving targets that may behave differently a few months from now. Third, as discussed in the introduction, contamination cannot be fully excluded, since most of the papers I study and their corrections predate the models' knowledge cutoffs. The \citet{Tirole1985} experiment, run with memory and web search disabled and almost concurrently to the publication of the correction, is the cleanest case on this score; for the remaining papers the picture is more mixed.

Taken together, the experiments suggest cautious optimism. Frontier AI is already a useful aid for scrutinizing economic theory, and is plausibly better than the current practice of referee reports written by humans who do not read proofs. But the headline question, whether AI can refute economic theory, is best answered as follows: not yet on
its own, but very well in the hands of an informed human who knows where to look.
	
\printbibliography
	
\end{document}